\documentclass[runningheads,a4paper]{llncs}

\usepackage{amssymb}
\usepackage{graphicx}

\usepackage{amsmath,xspace}
\usepackage{psfrag}

\usepackage{url}

\urldef{\emails}\path|macaule@clemson.edu, henning@vt.edu|

\newcommand{\keywords}[1]{\par\addvspace\baselineskip
\noindent\keywordname\enspace\ignorespaces#1}

\newcommand{\F}{\mathbb{F}}

\newcommand{\R}{\mathbb{R}}
\newcommand{\V}{\mathcal{V}}
\newcommand{\Z}{\mathbb{Z}}
\newcommand{\DG}{\mathcal{DG}}
\def\sds{{SDS}\xspace}

\def\<{\langle}
\def\>{\rangle}
\def\e{\vec{e}}
\def\v{\vec{v}}
\def\z{\vec{z}}
\DeclareMathOperator{\Acyc}{Acyc}
\DeclareMathOperator{\Aut}{Aut}
\DeclareMathOperator{\C}{C}
\DeclareMathOperator{\ECA}{ECA}
\DeclareMathOperator{\Per}{Per}
\DeclareMathOperator{\SL}{SL}
\DeclareMathOperator{\GL}{GL}
\def\simkappa{\!\!\sim_{\kappa}}
\def\tsimkappa{\sim_{\kappa}}

\def\tsimbarkappa{\sim_{\bar\kappa}}

\begin{document}

\mainmatter  

\title{Coxeter Groups and\\ Asynchronous Cellular Automata}
\titlerunning{Coxeter Groups and ACAs}

\author{Matthew Macauley\thanks{Department of Mathematical Sciences,
Clemson University.}  \and Henning S.~Mortveit\thanks{Department of
Mathematics and NDSSL/VBI, Virginia Tech.}}
\authorrunning{M.~Macauley and H.~S.~Mortveit}
\institute{}

\toctitle{Lecture Notes in Computer Science}
\tocauthor{Authors' Instructions}
\maketitle

\begin{abstract}
  The dynamics group of an asynchronous cellular automaton (ACA)
  relates properties of its long term dynamics to the structure of
  Coxeter groups.
  The key mathematical feature connecting these diverse fields is
  involutions. Group-theoretic results in the latter domain may lead
  to insight about the dynamics in the former, and vice-versa. In this
  article, we highlight some central themes and common structures, and
  discuss novel approaches to some open and open-ended problems. We
  introduce the state automaton of an ACA, and show how the root
  automaton of a Coxeter group is essentially part of the state
  automaton of a related ACA.
\end{abstract}

\section{Introduction}

An asynchronous cellular automaton (ACA) is defined in the same manner
as a classical cellular automaton (CA) in all aspects except the
evaluation mechanism. As the name suggests, the maps associated to the
vertices (or nodes) are applied synchronously for a CA, and
asynchronously for an ACA. In general, there are many ways that one
can apply maps asynchronously. For example, one may select a vertex at
random according to some probability distribution, apply the
corresponding map, and repeat this procedure. Alternatively, one may
select a fixed permutation over the vertices and apply the maps in the
sequence specified by this permutation. This permutation evaluation
process would correspond to increasing the time by one unit, and would
be applied repeatedly to generate the system dynamics. An important
aspect of having a fixed permutation update sequence is that one
obtains a dynamical system. This is not necessarily the case in the
more general situation, such as when the individual states are updated
at random.

The analysis of CAs and ACAs does not have the support that the study
of ODEs has from established fields such as analysis and differential
geometry. As such, a key goal of CA/ACA research is to make
connections to existing mathematical theory. We will consider the
class of \emph{$\pi$-independent} ACAs -- those whose periodic points
(as a set) are independent of the permutation update sequence. While
this may seem to be a rather exotic property, we have shown that
roughly~40\% of the elementary CA rules give rise to $\pi$-independent
ACAs~\cite{Macauley:08a}.
Given a $\pi$-independent ACA, one can define its \emph{dynamics
group}. This permutation group on the set of periodic points is a
quotient of a Coxeter group, and it captures the possible long-term
dynamics that one can generate by suitable choices of update
sequence. Its structure can answer questions about the existence and
non-existence of periodic orbits of given sizes.

In this paper, we will revisit the notions of Coxeter systems and
sequential dynamical systems (SDSs). An SDS is a generalization of an
ACA (assuming a fixed
update sequence) where the underlying graph is arbitrary, and is not
limited to being a regular lattice or circle (i.e., a one-dimensional
torus). We will show how the word problem for Coxeter groups is
related to functional equivalence of SDS maps. This forms the basis
for our next result, on how conjugation of Coxeter elements
corresponds to cycle equivalence of SDS maps, and additionally, how
this extends from conjugacy classes to spectral classes. After
defining dynamics groups and showing how they arise as quotients of
Coxeter groups, we show how key features of mathematical objects in
both the fields of SDSs and Coxeter groups are encoded by finite (or
infinite) state automata. We illustrate this by explicit examples,
and then close with a table summarizing these connections.


\section{Background}

A \emph{Coxeter system} is a pair $(W,S)$ consisting of a group $W$
generated by a set $S=\{s_1,\dots,s_n\}$ of involutions given by the
following presentation
\[
W=\<s_1,\dots,s_n\mid s_i^2=1,\;\;(s_is_j)^{m(s_i,s_j)}=1\>\,,
\]
where $m(s_i,s_j)\geq 2$ for $i\neq j$. Let $S^*$ be the free monoid
over $S$, and for each integer $m\geq 0$ and distinct generators
$s,t\in S$, define
\[
\<s,t\>_m=\underbrace{stst\cdots}_m\in S^*\,.
\]
The relation $\<s,t\>_{m(s,t)}=\<t,s\>_{m(s,t)}$ is called a
\emph{braid relation}, and is additionally called a \emph{short braid
relation} if $m(s,t)=2$. Note that $s$ and $t$ commute if and only if
$m(s,t)=2$. A Coxeter system can be described uniquely by its
\emph{Coxeter graph} $\Gamma$, which has vertex set $V=\{1,\dots,n\}$
and an edge $\{i,j\}$ for each non-commuting pair of generators
$\{s_i,s_j\}$, with edge label $m(s_i,s_j)$.

\medskip

Switching to ACAs and SDSs, let $\Gamma$ be an undirected graph
(called the \emph{base graph} or \emph{dependency graph}) with vertex
set $V=\{1,\dots,n\}$. We equip each vertex $i$ with a state $x_i\in
K$ where $K$ is a set called the \emph{state space}, and a
\emph{vertex function} $f_i$ that maps (or updates) $x_i(t)$ to
$x_i(t+1)$ based on the states of its neighbors (itself included).
Unless explicitly stated otherwise, we will assume that
$K=\F_2=\{0,1\}$, which is the most commonly used state space in
cellular automata research. If the vertex functions are applied
asynchronously, it is convenient to encode $f_i$ as a
\emph{$\Gamma$-local function} $F_i\colon K^n\to K^n$ defined by
\[
F_i(x_1,\dots,x_n) =
   (x_1,\dots,x_{i-1},f_i(x_1,\dots,x_n),x_{i+1},\dots,x_n)\;.
\]
If $f_i$ does not depend on all $n$ states, it may be convenient to
omit the fictitious variables. Given a sequence of local functions and
a word $w=w_1w_2\dots w_m\in V^*$ called the \emph{update sequence},
the {SDS map} $F_w$ is the composition of the local functions in the
order prescribed by $w$, i.e.,
\[
F_w \colon K^n\longrightarrow K^n\,,\qquad F_w  = F_{w_m}\circ
F_{w_{m-1}}\circ\cdots\circ F_{w_2}\circ F_{w_1}\,.
\]
SDSs represent a generalization of ACAs, which are usually defined
over a regular grid, such as $\Z$ or $\Z_n$. The following example
illustrates some SDS concepts -- see~\cite{Mortveit:07} for a more
complete treatment.

\begin{example}
\label{ex:one}
We take $\Gamma = \mathrm{Circ}_4$ as base graph (see
Figure~\ref{fig:nor_circ4}) and use $K = \{0,1\}$ as the state
space. Also, we take all vertex functions to be Boolean
$\mathrm{nor}$-functions given by $\mathrm{nor}\colon K^3\to K$ where
$\mathrm{nor}(x,y,z)$ equals $1$ if $x = y = z = 0$ and $0$
otherwise. In this case we have, for example, $F_1(x_1,x_2,x_3,x_4) =
(\mathrm{nor}(x_4,x_1,x_2), x_2, x_3, x_4)$. Using the update sequence
$\pi=1234$, we obtain
\begin{align*}
F_1(0,0,0,0) &= (1,0,0,0) \\
F_2 \circ F_1(0,0,0,0) &= (1,0,0,0) \\
F_3 \circ F_2 \circ F_1(0,0,0,0) &= (1,0,1,0) \\
F_4 \circ F_3 \circ F_2 \circ F_1(0,0,0,0) &= (1,0,1,0) \;,
\end{align*}
and thus $F_{\pi}(0,0,0,0) = (1,0,1,0)$. The phase space of the map
$F_{\pi}$ is the directed graph containing all global state transitions
and is shown in Figure~\ref{fig:nor_circ4}.
\begin{figure}[ht]
\centerline{
\raise20pt\hbox{\includegraphics[width=0.15\textwidth]{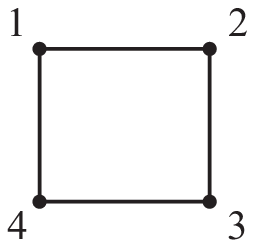}}
\qquad\qquad
\includegraphics[width=0.45\textwidth]{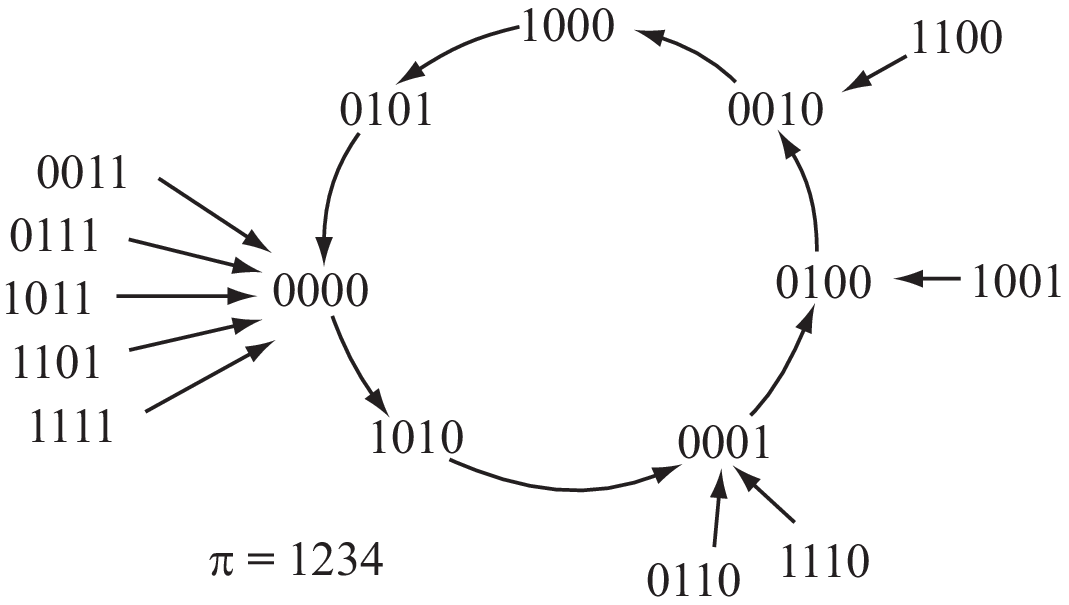}
}
\caption{The base graph $\Gamma=\mathrm{Circ}_4$ and the phase space
  of $F_{\pi}$ from Example~\ref{ex:one}.}
\label{fig:nor_circ4}
\end{figure}
\end{example}

\section{The word problem}

A fundamental question given any finitely presented group $\<S\mid
R\>$, is when do two words
\[
w=w_1w_2\cdots w_m\,,\quad\mbox{and}\quad
w'=w'_1w'_2\cdots w'_k
\]
in $S^*$ yield the same group element? This is the \emph{word
  problem}, and it is in general undecidable. However, there are many
  classes of groups for which the word problem is solvable. A classic
  result in Coxeter groups, known as \emph{Matsumoto's theorem}
  \cite[Theorem 1.2.2]{Geck:00}, says that any two reduced expressions
  for the same element differ by braid relations. Matsumoto's theorem
  provides an algorithmic solution to the word problem for Coxeter
  groups.

There is an analog of the word problem for SDSs. Specifically, given
two update sequences $w,w'\in V^*$, when are the corresponding SDS
maps
\[
F_w=F_{w_m}\circ F_{w_{m-1}}\circ\cdots\circ F_{w_2}\circ
F_{w_1}\,,\qquad F_{w'}=F_{w'_k}\circ F_{w'_{k-1}}\circ\cdots\circ
F_{w'_2}\circ F_{w'_1}\,
\]
equal as functions, or equivalently, when do they have identical phase
spaces? This is clearly solvable because there are only finitely many
functions $\F_2^n\to\F_2^n$. However, it would be desirable to solve
this problem algorithmically for general SDSs, without resorting to
checking the image of all $2^n$ global states.

\section{Equivalences on Dynamics and Acyclic Orientations}

In this section, we show how topological conjugation of SDS maps
corresponds to conjugation of elements in a Coxeter group, and how
this connection leads to a coarser equivalence relation when the graph
$\Gamma$ has non-trivial symmetries. \emph{Acyclic orientations} are
mathematically convenient to capture several types of equivalences on
permutation SDS maps, as well as on Coxeter elements in Coxeter
groups. A \emph{Coxeter element} is the product of the generators of
$S$ in some order. Every Coxeter element defines a partial ordering on
$S$, which we can represent by an acyclic orientation of $\Gamma$.
Specifically, for a Coxeter element $c$, define the orientation
$(\Gamma,c)$ so that edge $\{i,j\}$ is oriented $(i,j)$ if $s_i$
appears before $s_j$ in $c$. It is easy to show that this is
well-defined, and that it defines a bijection between the set
$\Acyc(\Gamma)$ of acyclic orientations of $\Gamma$ and the set
$\C(W)$ of Coxeter elements of $W$.

Next, consider conjugating a Coxeter element $c=s_{x_1}\cdots s_{x_n}$
by the initial letter $s=s_{x_1}$, which results in a cyclic shift of
the word:
\[
scs=s_{x_1}(s_{x_1}s_{x_2}\cdots s_{x_n})s_{x_1} =s_{x_2}s_{x_3}\cdots
s_{x_n}s_{x_1}\,.
\]
The corresponding acyclic orientations $(\Gamma,c)$ and $(\Gamma,scs)$
differ by converting the source vertex of $(\Gamma,c)$ into a
sink. This source-to-sink conversion generates an equivalence relation
$\tsimkappa$ on $\Acyc(\Gamma)$, and it was recently proven
(see~\cite{Eriksson:09}) that $(\Gamma,c)\tsimkappa(\Gamma,c')$ if and
only if $c$ and $c'$ are conjugate. (Note that the ``if'' direction is
obvious; the ``only if'' direction is difficult).

\medskip

Turning to SDSs, let $S_n\subset V^*$ be the set of words where each
vertex appears precisely once, which we may identify with the
permutations of $V$. Each permutation $\pi\in S_n$ defines a partial
ordering on $V$, and there is a natural map from $\Acyc(\Gamma)$ to
the set of permutation SDS maps ($\pi$ is mapped to $F_\pi$). Two
finite dynamical systems $\phi,\psi\colon K^n\to K^n$ are said to be
\emph{cycle equivalent} if for some bijection $h\colon K^n\to K^n$ we
have $ \psi|_{\Per(\psi)}\circ h=h\circ\phi|_{\Per(\phi)}\;, $ where
$\Per(\phi)$ denotes the set of periodic states of $\phi$.  The
following result provides the connection between $\kappa$-equivalence
of acyclic orientations and cycle equivalence of permutation SDS maps.
\begin{theorem}[\cite{Macauley:09a}]
  If $(\Gamma,\pi)\tsimkappa(\Gamma,\sigma)$, then $F_\pi$ and
  $F_\sigma$ are cycle equivalent.
\end{theorem}

\medskip

If the automorphism group $\Aut(\Gamma)$ is non-trivial, we can say
even more. The group $\Aut(\Gamma)$ acts on $\Aut(\Gamma)/\simkappa$
by $\gamma\cdot[(\Gamma,\pi)] = [(\Gamma,\gamma\pi)]$, which gives
rise to the equivalence relation $\tsimbarkappa$ on
$\Aut(\Gamma)/\simkappa$. This coarser equivalence relation also has
an interpretation in the settings of both Coxeter groups and SDSs.

If $(W,S)$ is a Coxeter system with $|S|=n$, let $\V$ be an
$n$-dimensional real vector space with basis
$\{\vec\alpha_1,\dots,\vec\alpha_n\}$. Put a symmetric bilinear form
$B$ on $\V$, defined by
$B(\vec\alpha_i,\vec\alpha_j)=-\cos\bigl(\pi/m(s_i,s_j)\bigr)$.
The group $W$ acts on $\V$ by
\begin{equation}
  \label{eq:action}
  s_i\colon\v\mapsto\v-2B(\v,\vec\alpha_i)\vec\alpha_i\,,
\end{equation}
and the set of elements $\Phi=\{w\vec\alpha_i\mid w\in
W\,,i=1,\dots,n\}$ are called \emph{roots}. This action is faithful
and preserves the bilinear form $B$. Geometrically, the root $s_i\v$
is the reflection of $\v$ across the hyperplane $\vec\alpha_i^\perp$,
and so there is a representation $\rho\colon W\to\GL(\V)$, defined on
the generators by
\begin{equation}
  \label{eq:representation}
  \rho\colon s_i\longmapsto\big(\v\stackrel{F_i}{\mapsto}
  \v-2B(\v,\vec\alpha_i)\vec{\alpha}_i\big)\,,
\end{equation}
called the \emph{standard geometric representation} of $W$
(see~\cite{Bjorner:05,Humphreys:90}). This allows us to view elements in
$W$ as matrices, and hence we can speak of the characteristic
polynomial of any given $w\in W$.

Now, if $(\Gamma,c)$ and $(\Gamma,c')$ differ by some
$\gamma\in\Aut(\Gamma)$, then $\rho(c)$ and $\rho(c')$ are similar as
linear transformations.  Specifically, they are conjugate in $\GL(\V)$
by the permutation matrix $P_\gamma$ of $\gamma$. In this case, we say
that $c$ and $c'$ have the same spectral class, because $\rho(c)$ and
$\rho(c')$ have the same multiset of eigenvalues. Clearly, this is a
weaker condition than conjugacy, and so all Coxeter elements in the
same $\bar\kappa$-equivalence class have the same spectral class.

\medskip

Similarly, in the context of SDSs, if $(\Gamma,\pi) \tsimbarkappa
(\Gamma,\sigma)$, then the SDS maps $F_\pi$ and $F_\sigma$ are cycle
equivalent, due to the following argument. If $\gamma\in
\Aut(\Gamma)$, then the permutations $\pi$ and $\gamma\pi$ give
topologically conjugate SDS maps, $F_{\pi}$ and
$F_{\gamma\pi}$. Strictly speaking, this requires the maps $f_i$ to be
$\Aut(\Gamma)$-invariant (see~\cite{Macauley:09a}), a condition which
is frequently satisfied in practice, such as when all vertices of the
same degree share the same symmetric function (e.g., logical AND, OR,
Majority, Parity, threshold functions, etc.).
Since topologically conjugate maps are cycle equivalent, our statement
follows.

It is worth mentioning the role of the Tutte polynomial
here~\cite{Tutte:54}. The Tutte polynomial of a graph $\Gamma$ is a
2-variable polynomial $T_\Gamma(x,y)$ that satisfies a recurrence
under edge deletion and contraction, and plays a central role in graph
theory. Many graph counting problems are simply the evaluation of the
Tutte polynomial at some $(x_0,y_0)\in\Z\times\Z$. For example,
$|\Acyc(\Gamma)| = T_\Gamma(2,0)$ and
$|\Acyc(\Gamma)/\simkappa\!\!|=T_\Gamma(1,0)$. Thus, $T_\Gamma(2,0)$
counts the number of Coxeter elements in the Coxeter group with
Coxeter graph $\Gamma$, and it bounds the number of permutation SDS
maps in an SDS with dependency graph $\Gamma$. This bound is known to
be sharp for certain classes of functions~\cite{Mortveit:07}.
Similarly, $T_\Gamma(1,0)$ counts the number of conjugacy classes of
Coxeter elements (see~\cite{Eriksson:09,Macauley:08b}), and it bounds
the number of cycle equivalence classes of SDS maps
(see~\cite{Macauley:09a}).

\section{Groups}

A sequence $F=(F_1,\dots,F_n)$ of local functions is
\emph{$\pi$-independent} if $\Per(F_\pi)=\Per(F_\sigma)$ for all
$\pi,\sigma\in S_n$. Note that this is an equality of sets; we do not
assume anything about the organization of the respective periodic
points into periodic orbits. In this case, each $F_i$ permutes the
periodic points, and these permutations generate the \emph{dynamics
group} of $F$, denoted $\DG(F)$. Let $F_i^*$ denote the restriction of
$F_i$ to $\Per(F_\pi)$. Because $F_i$ only changes the $i^{\rm th}$
coordinate of a state, and since we assume that $K=\F_2$, $F_i^*\circ
F_i^*$ is the identity function on $\Per(F_\pi)$. If we define
$m_{ij}:=|F^*_i\circ F^*_j|$, then there is a surjection
\begin{equation}
  \label{eq:quotient}
  \<s_1,\dots,s_n\mid
  s_i^2=1,\;(s_is_j)^{m_{ij}}=1\>\longrightarrow\DG(F)\,,
\end{equation}
showing that dynamics groups are quotients of Coxeter groups. The
particular homomorphism is determined by adding relations to the
presentation of the Coxeter group, and these relations arise because
the state space is $\F_2$. Thus, dynamics groups are in a sense
``reflection groups over $\F_2$.'' An open-ended research problem is
to give an efficient presentation of the dynamics groups of an SDS
based on the functions, i.e., to determine these extra relations.

When the base graph $\Gamma$ of an SDS is the circular graph $\Z_n$,
and the local functions are all identical, the resulting SDS is an
\emph{elementary ACA}. Each local function $F_i$ takes
$\{x_{i-1},x_i,x_{i+1}\}$ as input, and is completely described by the
following rule table
\[
\begin{array}{c||c|c|c|c|c|c|c|c}
  x_{i-1}x_ix_{i+1} & 111 & 110 & 101 & 100 & 011 & 010 & 001 & 000
  \\ \hline f_i(x_{i-1},x_i,x_{i+1}) & a_7 & a_6 & a_5 & a_4 & a_3
  & a_2 & a_1 & a_0
\end{array}
\]
Clearly, there are $2^{2^3}=256$ such choices of functions, which can
be indexed by $k=\sum a_i2^i\in\{0,\dots,255\}$. The corresponding
sequence of local functions is denoted
$\ECA_k$. In~\cite{Macauley:08a}, it was shown that $\ECA_k$ is
$\pi$-independent for precisely 104 values of $k$. Moreover, this
holds for all $n>3$. The dynamics groups of these 104 rules were
classified in~\cite{Macauley:10b}. Among some of the interesting
groups were $\DG(\ECA_{60})=\SL_n(\F_2)$ and $\DG(\ECA_k)=\Z_2^n$ for
$k\in\{28,29,51\}$. Moreover, other dynamics groups were found
computationally to be either the symmetric or alternating groups, with
the size depending on the $n^{\rm th}$ Fibonacci or Lucas number,
leading to a few conjectures.

\section{The root automaton}

\begin{figure}[ht]
\centerline{\includegraphics[width=0.5\textwidth]
  {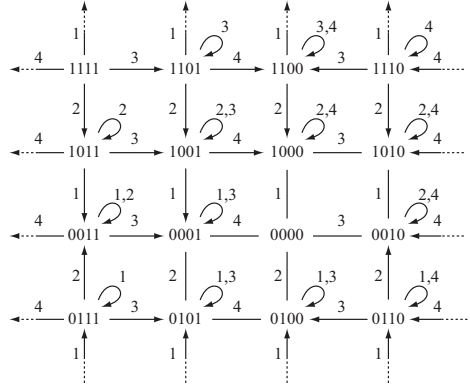}}
\caption{The state automaton $\Phi$ for the SDS in
  Example~\ref{ex:one}. Horizontal/vertical dashed lines and arrows
  indicate horizontal/vertical wrap-around, and arrowheads are omitted
  from the bidirectional edges for clarity.}
\label{fig:state-aut}
\end{figure}

The dynamics of all possible SDSs given a sequence of $\Gamma$-local
functions $F=(F_1,\dots,F_n)$ can be encoded by the \emph{state
automaton} of the sequence. This is a directed graph $\Phi$ with
vertex set $K^n$ -- the set of global system states, and directed
edges $(x,F_i(x))$ for each $x\in K^n$ and each $i\in V$. Label such
an edge with the index $i$ corresponding to its vertex function; see
Figure~\ref{fig:state-aut} for an example.

The image of a state $x\in K^n$ under an SDS map $F_\pi$, where
$\pi=\pi_1\pi_2\cdots\pi_n$, is represented on the state automaton by a
path in $\Phi$. Specifically, start at vertex $x$ and traverse the
path
\[
x\,,\quad F_{\pi_1}(x)\,,\quad
F_{\pi_2}F_{\pi_1}(x)\,,\quad\dots\quad,\, F_{\pi_n}\!\cdots
F_{\pi_2}F_{\pi_1}(x)=F_\pi(x)\,.
\]
The phase space of $F_\pi$ can be easily derived from the state
automaton -- it is the graph with vertex set $K^n$ and an edge
$(x,y)$ for every directed path from a state $x$ to $y$ that traverses
a path of edges labeled $\pi_1,\pi_2,\dots,\pi_n$. Note that if $F$ is
$\pi$-independent, then $\DG(F)$ acts on $\Per(F)$. In this case, all
of the edges within $\Per(F)$ are bidirectional, and so we may view
them as undirected.

This is the SDS analog of the action of $W$ on $\Phi\subset\V$, as
described in~\eqref{eq:action}. Since $\V$ is any $n$-dimensional
vector space, we can identify it with $\R^n$, and assume that the
basis elements are $\vec\alpha_i=\e_i$, the standard unit normal
vectors. This associates roots with vectors in $\R^n$, and we
partially order $\Phi$ by $\leq$ componentwise ($\z\preceq\z'$ iff
$z_i\leq z'_i$ for each $i$) to get the \emph{root poset}. It is
well-known that for every root, all non-zero entries have the same
sign, thus we have a notion of positive and negative roots, and the
root poset has a positive side $\Phi^+$ and a negative side, $\Phi^-$,
with $\Phi=\Phi^+\cup\Phi^-$.  The image of $s_i$ under the geometric
representation from~\eqref{eq:representation} is a linear map
$F_i\colon\R^n\to\R^n$, where
\begin{equation}
  \label{eq:F_i}
  F_i\colon (z_1,\dots,z_n)\longmapsto
  (z_1,\dots,z_{i-1},z_i+\sum_{j=1}^n 2\cos(\pi/m_{i,j})
  z_j,z_{i+1},\dots,z_n)\;.
\end{equation}
To summarize, $F_i$ changes the $i^{\rm th}$ entry of a vector by
flipping its sign and then adding each neighboring state $z_j$
weighted by $2\cos(\pi/m_{ij})$.

In 1993, Brink and Howlett proved that Coxeter groups are
automatic~\cite{Brink:93}, and soon after, H.~Eriksson developed the
\emph{root automaton}~\cite{ErikssonH:94}. The root automaton has
vertex set $\Phi$ and edge set $\{(\z,s_i\z)\mid\z\in\Phi,\,s_i\in
S\}$. For convenience, label each edge $(\z,s_i\z)$ with the
corresponding generator $s_i$. It is clear that upon disregarding
loops and edge orientations (all edges are bidirectional anyways), we
are left with the Hasse diagram of the root poset. We represent a word
$w=s_{x_1}s_{x_2}\cdots s_{x_m}$ in the root automaton by starting at
the unit vector $\e_{x_1}\in\Phi^+$ and traversing the edges labeled
$s_{x_2},s_{x_3},\dots,s_{x_m}$ in sequence. Denote the root reached
in the root poset upon performing these steps by $\vec{r}(W,w)$. The
sequence
\[
\e_{x_1}=\vec{r}(W,s_{x_1})\,,\quad\vec{r}(W,s_{x_1}s_{x_2})
\,,\quad\dots\quad,\,\vec{r}(W,s_{x_1}s_{x_2}\cdots s_{x_m})=\vec{r}(W,w)\,,
\]
is called the \emph{root sequence} of $w$. If
$\vec{r}(W,s_{x_1}s_{x_2}\cdots s_{x_i})$ is the first negative root
in the root sequence for $w$, then a shorter expression for $w$ can be
obtained by removing $s_{x_1}$ and $s_{x_i}$. By the \emph{exchange
property} of Coxeter groups~(see \cite{Bjorner:05,Humphreys:90}),
every word $w\in S^*$ can be made into a reduced expression by
iteratively removing pairs of letters in this manner. Thus, the root
automaton can algorithmically detect reduced words.

We conclude with an example that illustrates these concepts, and shows
how the root automaton of a Coxeter group is essentially a connected
component of the state automaton of an sequential dynamical system
with state space $K=\R$.

\begin{example}
Let $W=H_4$, which has Coxeter graph as shown in Figure~\ref{fig:h4}, and
presentation (using $a,b,c,d$ instead of $s_1,s_2,s_3,s_4$):
\[
H_4=\<a,b,c,d\mid a^2,\,b^2,\,c^2,\,d^2,\,
(ab)^5,\,(bc)^3,\,(cd)^3,\,(ac)^2,\,(ad)^2,\,(bd)^2\>\,.
\]
\begin{figure}[ht]
  \psfrag{1}{$_a$}\psfrag{2}{$_b$}\psfrag{3}{$_c$}\psfrag{4}{$_d$}
  \psfrag{5}{$5$}
  \centerline{\includegraphics[width=0.4\textwidth]{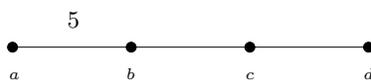}}
  \caption{The Coxeter graph $\Gamma$ of the group $W=H_4$. As is
  customary, edge labels of $3$ are suppressed.}
  \label{fig:h4}
\end{figure}
It is well-known (see~\cite{Humphreys:90}) that $H_4$ is a finite
group of order $14400$, and is the isometry group of the $120$-cell
and its dual, the $600$-cell, two of the six regular
$4$-polytopes. Thus, the root poset $\Phi$ consists of $14400$
roots. A portion of the root automaton is shown in
Figure~\ref{fig:h4-automaton}.
\begin{figure}[ht]
  \psfrag{(-1000)}{$_{(-1,0,0,0)}$}
  \psfrag{(0-100)}{$_{(0,-1,0,0)}$}
  \psfrag{(00-10)}{$_{(0,0,-1,0)}$}
  \psfrag{(000-1)}{$_{(0,0,0,-1)}$}
  \psfrag{(1000)}{$_{(1,0,0,0)}$}
  \psfrag{(0100)}{$_{(0,1,0,0)}$}
  \psfrag{(0010)}{$_{(0,0,1,0)}$}
  \psfrag{(0001)}{$_{(0,0,0,1)}$}
  \psfrag{(1p00)}{$_{(1,\phi,0,0)}$}
  \psfrag{(p100)}{$_{(\phi,1,0,0)}$}
  \psfrag{(0110)}{$_{(0,1,1,0)}$}
  \psfrag{(0011)}{$_{(0,0,1,1)}$}
  \psfrag{(1pp0)}{$_{(1,\phi,\phi,0)}$}
  \psfrag{(pp00)}{$_{(\phi,\phi,0,0)}$}
  \psfrag{(p110)}{$_{(\phi,1,1,0)}$}
  \psfrag{(0111)}{$_{(0,1,1,1)}$}
  \psfrag{(1ppp)}{$_{(1,\phi,\phi,\phi)}$}
  \psfrag{(ppp0)}{$_{(\phi,\phi,\phi,0)}$}
  \psfrag{(pq10)}{$_{(\phi,\phi^2,1,0)}$}
  \psfrag{(p111)}{$_{(\phi,1,1,1)}$}
  \psfrag{a}{$_a$}\psfrag{b}{$_b$}\psfrag{c}{$_c$}\psfrag{d}{$_d$}
  \psfrag{a,b}{$_{a,b}$}\psfrag{b,c}{$_{b,c}$}\psfrag{c,d}{$_{c,d}$}
  \centerline{\includegraphics{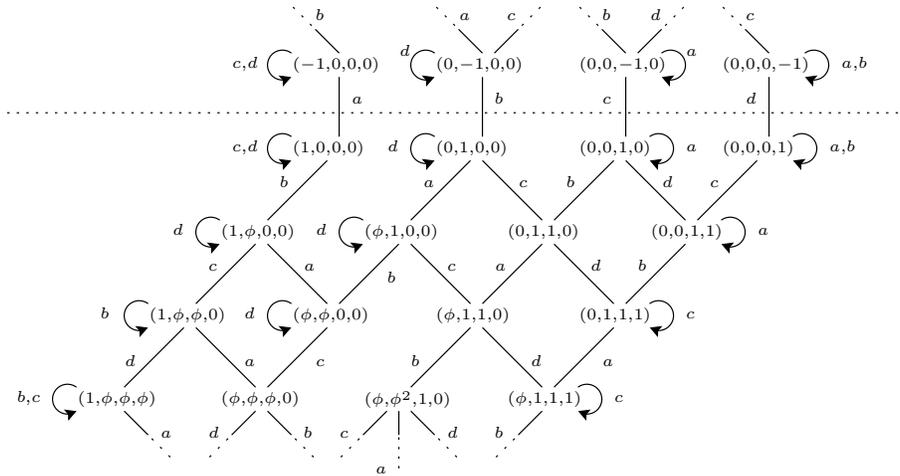}}
  \caption{Part of the root automaton of the group $W=H_4$. Here,
    $\phi=2\cos(\pi/5)$, the golden ratio. The dotted line separates the
    positive roots from the negative roots.}
  \label{fig:h4-automaton}
\end{figure}
Recall that the root automaton is built on top of the root poset --
stripping away the self-loops and edge labels leaves the Hasse diagram
of $\Phi$. The dotted-line in Figure~\ref{fig:h4-automaton} shows the
boundary between the positive roots $\Phi^+$ and negative roots
$\Phi^-$. The non-loop edges of the root automaton are all
bidirectional -- arrowheads are omitted for clarity.

Consider the word $w=abdcabacbca\in H_4$. Starting at $\e_a=(1,0,0,0)$
(see Figure~\ref{fig:h4-automaton}), and traversing the edges labeled
$b,d,c,a,b,a,c,b,c,a$ in sequence, we see that the first negative root
in the root sequence of $w$ is $\vec{r}(W,abdcabacbc)$. Therefore,
removing the first instance of $a$ and the last instance of $c$ from
$w$ results in $bdcabacba$, a shorter expression for $w$. It
is easily checked that no matter where we begin in $bdcabacba$, the
corresponding path in the root automaton consists of only positive
roots. Therefore, $bdcabacba$ is a reduced word in $H_4$.
\end{example}

\section{Summary}

This paper presented a collection of results connecting properties of
Coxeter groups and properties of the dynamics of ACAs/SDSs (see
Table~\ref{tbl:summary}).
\begin{table}[ht]
\centerline{
  \begin{tabular}{p{0.60in}ccc|cc}
    & & Coxeter groups &$\quad$&$\quad$& Sequential dynamical systems
    \\ \hline
    &&& \\
    Graph $\Gamma$ & $\longleftrightarrow\quad$ & Coxeter
    graph &$\quad$&$\quad$& Dependency graph \\
    &&& \\
    $\Acyc(\Gamma)$ & $\longleftrightarrow\quad$ & Coxeter elements
    &$\quad$&$\quad$& Permutation SDS maps \\
    && $w=s_{\pi(1)}s_{\pi(2)}\cdots s_{\pi(n)}$ &$\quad$&$\quad$&
    $F_\pi=F_{\pi(n)}\circ\cdots\circ F_{\pi(2)}\circ
    F_{\pi(1)}$. \\ &&& \\
    $\kappa$-equiv. & $\longleftrightarrow\quad$ & Conjugacy classes
    &$\quad$&$\quad$& Cycle-equivalence classes \\
    && of Coxeter elements &$\quad$&$\quad$& of \sds maps \\ &&& \\
    $\bar\kappa$-equiv. & $\longleftrightarrow\quad$ &
    Spectral classes &$\quad$&$\quad$& Cycle-equivalence classes \\
    && of Coxeter elements &$\quad$&$\quad$& of \sds maps (coarser) \\
    &&& \\
    $\Phi$ & $\longleftrightarrow\quad$ & Root poset / automaton
    &$\quad$&$\quad$& State automaton \\
    &&&
  \end{tabular}
}
  \caption{Summary of the connections between Coxeter groups and
    SDSs.\label{tbl:summary}}
\end{table}
These newly established connections provide possible avenues for ACA
research. In a larger setting, we hope that our example linking
properties of asynchronous, finite dynamical systems and group theory
can provide inspiration for other approaches seeking to better
understand the dynamics of ACAs through the use of existing
mathematical structures and theory.

\subsubsection*{Acknowledgments.}
The authors thank the Network Dynamics and Simulation Science
Laboratory at the Virginia Bioinformatics Institute of Virginia Tech
and Ilya Shmulevich's research group at the Institute for Systems
Biology for their support.

\end{document}